\documentclass{emulateapj}

\usepackage{graphicx}
\usepackage{txfonts}
\usepackage{amssymb}
\usepackage{color}
\usepackage{subfigure}
\usepackage{natbib}
\bibpunct{(}{)}{;}{a}{}{,} 
\newcommand{\nc}{\newcommand}
\nc{\lsun}{\ensuremath{\mathrm{L}_\odot}}
\nc{\Msun}{\ensuremath{\mathrm{M}_\odot}}
\nc{\tex}{\ensuremath{\mathrm{T}_{\rm ex}}}
\nc{\cthree}{C$_3$}
\nc{\cthreehtwo}{$c$-C$_3$H$_2$}
\nc{\kms}{\mbox{km\,s$^{-1}$}}
\nc{\Kkms}{\mbox{K\,km\,s$^{-1}$}}
\def\ptsec{$''\mskip-7.6mu.\,$}

\nc{\Trot}{$T_{\rm rot}$}%
\nc{\Ntot}{$N(C_3)$}%
\nc{\Tc}{$T_{\rm c}$}%
\nc{\Tdust}{$T_{\rm dust}$}%
\nc{\Tex}{$T_{\rm ex}$}%
\nc{\Tkin}{$T_{\rm kin}$}%
\nc{\Tmax}{$T_{\rm max}$}%
\nc{\cmcub}{\mbox{cm$^{-3}$}}
\nc{\cmsq}{\mbox{cm$^{-2}$}}

\newcommand{\CII}{[C {\sc ii}]}
\newcommand{\OI}{[O {\sc i}]}
\newcommand{\emm}[1]{\ensuremath{#1}}   

\renewcommand{\deg}{\emm{^\circ{}}}

\shorttitle{\OI\ in the HD\,50138 circumstellar disk}
\shortauthors{Sandell et al.}

\begin{document} 

\title{Velocity resolved \OI\ 63 $\mu$m emission in the HD\,50138 circumstellar disk  \thanks{Based in part on observations made with
  ESO Telescopes at the La Silla Paranal Observatory under programme ID 384.D-0613(A)} }

   \author{G\"oran Sandell}
    \affil{University of Hawaii, Institute for Astronomy - Hilo,  640 N. Aohoku Place, Hilo, HI 96720, USA}
     
    \email{gsandell@hawaii.edu} \label{IfA}

    \author{C. Salyk}
     \affil{Vassar College, 124 Raymond Avenue, Poughkeepsie, NY 12604, USA}

   \author{ M. van den Ancker}
   \affil{ European Southern Observatory, Karl-Schwarzschild-Str. 2, 85748 Garching bei M\"unchen, Germany}
    
    \author{W.-J. de Wit}
    \affil{European Southern Observatory, Alonso de C\'ordova 3107, Casilla 19001, Santiago, Chile}
      
      \author{E. Chambers} 
      \affil{SOFIA Science Center, NASA Ames Research Center\\
              UNIVERSITIES SPACE RESEARCH ASSOCIATION\\
              MS 232-12, Building N232, PO Box 1, Moffett Field, CA 94035-0001, USA}
      
    \author{R. G\"usten}
    \affil{ Max Planck Institut f\"ur Radioastronomie, Auf dem H\"ugel 69, 53121 Bonn, Germany}
    
     \author{H. Wiesemeyer}
     \affil{ Max Planck Institut f\"ur Radioastronomie, Auf dem H\"ugel 69, 53121 Bonn, Germany}
                 
       \author{H. Richter}
       \affil{DLR e.V., Institut f\"ur Optische Sensorsysteme, Rutherfordstr. 2, 12489 Berlin, Germany}
           

  \begin{abstract}

 HD\,50138 is one of the brightest B[e] stars at a distance of $\sim$ 380 pc
 with strong infrared excess. The star was observed in \OI\ 63 $\mu$m and \CII\
 158 $\mu$m with high velocity resolution with upGREAT on SOFIA. The velocity
 resolved \OI\ emission provides evidence for a large gas-disk, $\sim$ 760 au in
 size, around HD\,50138. Whereas previous interferometric observations give
 strong evidence for a hot gas and dust disk in Keplerian rotation, our
 observations are the first to provide unambiguous evidence for a large warm
 disk around the star. {\it Herschel}/PACS observations showed that the \CII\
 emission is extended, therefore the  \CII\ emission most likely originates in
 an ionized gas shell created by a past outflow event. We confirm the isolated
 nature of HD\,50138. It is far from any star forming region and has low proper
 motion. Neither is there any sign of a remnant cloud from which it could have
 formed. The extended disk around the star appears carbon poor. It shows OH and
 \OI\ emission, but no CO. The CO abundance appears to be at least an order of
 magnitude lower than that of OH. Furthermore  $^{13}$CO is enriched by more than a factor of five, confirming that the star is not a
 Herbig Be star. Finally we note that our high spectral resolution \OI\ and
 \CII\ observations provide a very accurate heliocentric velocity of the star,
 40.8 $\pm$ 0.2 km~s$^{-1}$.
 
 \end{abstract}
  


\section{Introduction}

HD\,50138 (MWC\,158) is a bright, isolated B star with strong emission lines. It
was found to have variable hydrogen lines in an objective-prism survey for
bright-line B stars by \citet{Merrill25}. The spectral variability was later
confirmed by \citet{Merrill31} and \citet{Merrill33}, who suggested two periods
of variability, one long term period with a periodicity of about five years and
a shorter one of about 30 days. Another study by \citet{Doazan65} found the star
to have an expanding envelope with the speed of expansion changing on a
timescale of 50 days, although later studies have not been able to confirm these
short term periodicities.  Instead a recent paper \citep{Jerabkova16}, analyzing
spectra over the last  20 years  finds two long term periods, 8.2 $\pm$ 1.3 and 13.7 $\pm$ 2.7 years.
 They also find evidence for variability of the
order of 50 days, consistent with the results by \citet{Doazan65}. Some of these
variabilities have been explained as a result of an outburst that possibly took
place in 1978 -- 1979 \citep{Hutsemekers85} and by a shell phase in1990 -- 1991
\citep{Andrillat91} with yet another shell phase sometimes before 2007
\citep{Borges09}. The star has strong infrared excess \citep{Allen76} and
polarimetry and spectro-polarimetry suggest it is surrounded by a circumstellar
disk \citep{Vaidya94,Bjorkman98,Harrington07}.  Mid infrared interferometry
confirms the presence of an asymmetric disk in Keplerian rotation
\citep{Ellerbroek15} with a moderate inclination of 56\deg\ \citep{Borges11}.
\citet{Ellerbroek15} determine a spectral type B7 III in agreement with
\citet{Borges09}. The second {\it Gaia} release \citep{Gaia16, Gaia18} provides a
very accurate distance, 380 $\pm$ 9 pc. This is the distance we will
adopt in this paper.


B[e]-type stars, i.e.,stars that show the B[e] phenomenon, is a heterogenous
group of stars of different mass and different evolutionary status.
\citet{Lamers98} found that stars showing the B[e] phenomenon could be grouped
into five classes:B[e] supergiants, pre-main-sequence B[e]-type star or Herbig Ae/Be (HAEBE)
stars, compact planetary nebulae B[e]-type stars, symbiotic B[e]-type stars, and
unclassified B[e]-type stars. \citeauthor{Lamers98} summarize the criteria for stars
showing the B[e] phenomenon as follows. They all have: "strong Balmer emission
lines, low excitation permitted emission lines of predominantly low ionization
metals in the optical spectrum, e.g. \ion{Fe}{2}, forbidden emission lines of
[\ion{Fe}{2}] and \OI\ in the optical spectrum, and a strong near or
mid-infrared excess due to hot circumstellar dust."  \citet{Miroshnichenko07} argue
that most of the unclassified B[e] stars are FS CMa stars. FS CMa stars are believed to 
be close binary systems, which are currently going through, or have gone through a phase
of rapid mass transfer, resulting in mass loss and dust formation. Because the spectra of 
different types of B[e]-stars look the same, it is often very difficult to
determine whether a star is a pre-main-sequence, main,  or post main sequence
star. \citet{Kraus09} showed that if the CO bands are detected, they can be used to
obtain an age estimate for a star, because evolved stars should become enriched in $^{13}$CO.
This method has been successfully used by \citet{Liermann14}, who  showed that several stars in their sample were evolved because of
enhanced $^{13}$CO bandhead emission. 

High resolution emission line spectroscopy is a powerful tool to probe spatially  unresolved regions in circumstellar disks \citep[see e.g.][]{ Najita07,Brittain15}
Velocity resolved emission lines can provide spatial information about the gas if the disk is in Keplerian rotation and the inclination angle and  mass of the central star is
known. This is particularly important in the far infrared, where there are no interferometers and where the spatial resolution of large space based telescopes is
insufficient to resolve even the largest protoplanetary disks.

Even though HD\,50138 originally was classified as a classical B[e] star or as
being in transition from B[e] to Be \citep{Jaschek93}, the star shares many
characteristics with Herbig Be stars. \citet{Morrison95} and \citet{Grady96}
proposed that it is a pre-main-sequence (PMS) star, because it has a large infrared
excess, He {\sc I} and Si {\sc II} show inverse P Cygni profiles, i.e., evidence
for accretion, and the line profiles are consistent with a disk geometry, \citep[see
also][]{Pogodin97}. However, it lacks one of the important criteria for
Herbig Be stars, i.e., association with a starforming region \citep{Lee16}, making
it highly unlikely that HD\,50138 is a pre-main-sequence star and they argued that it is an FS CMa star.

We had included HD\,50138 in our small GREAT survey of \OI\ in circumstellar
disks, based on the strong \OI\ emission found in the {\it Herschel} Open Time (OT) key
project  DIGIT (Dust, Ice and Gas In Time) \citep{Fedele13}. However, since HD\,50138 does not appear to be a
pre-main-sequence star, and since the extended disk appears CO deficient, 
we are publishing our results of this star separately.

 
\section{Observations and archive data}

HD\,50138 ($\alpha$(2000.0) =$6^h51^m33.^s399$, $\delta$(2000.0) = $-$06\arcdeg\
57\arcmin\ 59\farcs5) was observed in \OI\ 63 $\mu$m with the upGREAT\footnote{The development of upGREAT (German REceiver for Astronomy at Terahertz frequencies) was financed by the participating
institutes, by the Federal Ministry of Economics and Technology via the German
Space Agency (DLR) under Grants 50 OK 1102, 50 OK 1103 and 50 OK 1104 and within
the Collaborative Research Centre 956, sub-projects D2 and D3, funded by the
Deutsche Forschungsgemeinschaft (DFG).} High
Frequency Array (HFA) during  GREAT  consortium time onboard the Stratospheric Observatory for Infrared Astronomy (SOFIA) on November 1, 
2016. The observations were done on a 71 minute leg at an altitude of  44,000
ft (13.6 km).  GREAT is a modular heterodyne instrument, with two channels, both
of which are used simultaneously. For a more complete description of the
instrument, see \citet{Heyminck12} and \citet{Risacher16}. Here we used the
recently commissioned HFA together with the low-frequency channel L2. HFA 
targets the \OI\  $^3{\rm P}_1 \to\ ^3{\rm P}_2$ transition at 4.74477749 THz,
while L2 was tuned  to the \CII\ $^2{\rm P}_{3/2} \to\  ^2{\rm P}_{1/2}$
transition at 1.9005369 THz in the upper sideband. The HFA is a hexagonal array
co-aligned around a central pixel, providing a 7 pixel array with the pixels
separated by two beam widths. The HFA cryostat is cooled with a closed cycle
pulse tube refrigerator and uses a novel quantum cascade laser as the local
oscillator. For a more complete description of the array,  see Risacher et al.
(2018, in prep) or \citet{Risacher16}, which describes the Low Frequency Array (LFA).
The LFA has similar design, except that it covers the 1.8 - 2.07 THz region and
uses orthogonal polarizations,  therefore providing 2 $\times$ 7 pixels
rather than 7 pixels. The main beam coupling efficiency, $\eta_{mb}$, was 0.69
for L2, and 0.65 for the central pixel of HFA. The half power beam width (HPBW)
for the HFA is $\sim$ 6\ptsec3, while the HPBW for L2 is 14\ptsec4 at 1.9
THz.  The boresight of the LFA array has an uncertainty of 1 - 2 \arcsec. The single-sideband system 
temperature was $\sim$ 2700 K for \CII\ and around 2800 -- 3200
K for \OI.

The backends for both channels are the last generation of fast Fourier transform
spectrometers (FFTS) \citep{Klein12}, with 4 GHz bandwidth and 16384 channels
providing a channel separation of 244.1 kHz (0.0385 km~s$^{-1}$ for \CII{}). The
data were reduced and calibrated by the GREAT team. The post processing was done
using the Continuum and Line Analysis Single-dish Software package
CLASS\footnote{CLASS is part of the Grenoble Image and Line Data Analysis
Software (GILDAS), which is provided and actively developed by IRAM, and is
available at http://www.iram.fr/IRAMFR/GILDAS} . We removed linear baselines,
threw away a few damaged spectra and coadded the spectra with rms weighting. The
final integration time (on+off) was 44 minutes for \CII, 38 minutes for \OI. The
final averages were resampled to 0.5 km~s$^{-1}$.

Long integration spectra of CO(3--2) and CO(4--3) were obtained with the 
First light APEX Submillimeter Heterodyne receiver FLASH$^+$ on the 
Atacama Pathfinder Experiment telescope (APEX) \citep{Gusten06} in Chile on
March 27, 2017. Both observations were done in dual beam switch mode with a
60\arcsec\ chop throw. The weather conditions were marginal for CO(4--3), $\tau$
= 1.06, system temperature, T$_{sys}$ $\sim$ 1800~K, but fine for CO(3--2),
$\tau$ $\sim$ 0.23, and T$_{sys}$ $\sim$ 270~K. The total integration time was
1.3 hrs and 2.6 hrs for CO(3--2) and CO(4--3), respectively. There is no sign of
any CO emission.
 

A near-infrared high spectral-resolution (R $\sim$ 50,000) spectrum of HD 50138,
covering the 2.276--2.326 $\mu$m wavelength range (containing the bandhead of CO
2-0 overtone emission) was taken with the VLT Cryogenic high-Resolution InfraRed
Echelle Spectrograph
(CRIRES\footnote{http://www.eso.org/sci/facilities/paranal/instruments/crires/},
\citep{Kaufl04}) on Nov. 5, 2009. Adaptive Optics were used to optimize the
signal-to-noise ratio and spatial resolution of the observations. The
observations were made with a slit width of 0\farcs4, with the slit oriented
along the parallactic angle. The spectrum  was reduced with the ESO
automatic pipeline, and corrected for telluric and heliocentric velocity.

We also analyze a spectrum obtained with NIRSPEC \citep{McLean98}, a
high-resolution echelle spectrograph on the Keck 2 telescope  on October 9,
2009. This M band spectrum of HD\,50138 was observed with a 0\farcs43 $\times$
24\arcsec\ slit, which provides a resolution, R A$\sim$ 25000 (FWHM $\sim$ 12.5
km~s$^{-1}$). The two spectral orders cover the wavelength range 4.65 -- 4.78
and 4.96 -- 5.1 $\mu$m. These wavelength ranges include the CO fundamental
($\upsilon = 1 \rightarrow 0$) rovibrational transitions R(0 - 1) and P(1 - 12,
30 - 40). The data were corrected for atmospheric absorption using standard
stars and wavelength corrected using telluric emission lines. For further
details on data reduction, see \citet{Salyk09}. \citet{Salyk11} used the data in
their study of molecular emission from protoplanetary disk, but did not show a
spectrum. 



We retrieved three photometry observations done with the Photoconductor 
Array Camera and Spectrometer (PACS) instrument on the {\it Herschel} Space
Observatory from the {\it Herschel} data archive. These were all done in the 70
$\mu$m and 160 $\mu$m bands (AOR-ID 1342228369, 1342228916, and 1342250822). All
three data sets are consistent within observational errors and give flux
densities of 7.50 $\pm$ 0.05 Jy, and 1.40 $\pm$ 0.04 Jy at 70 and 160 $\mu$m,
respectively.

\section{Results and Analysis}

\begin{figure}[t]
\includegraphics[angle=0,width=8.0cm,angle=0]{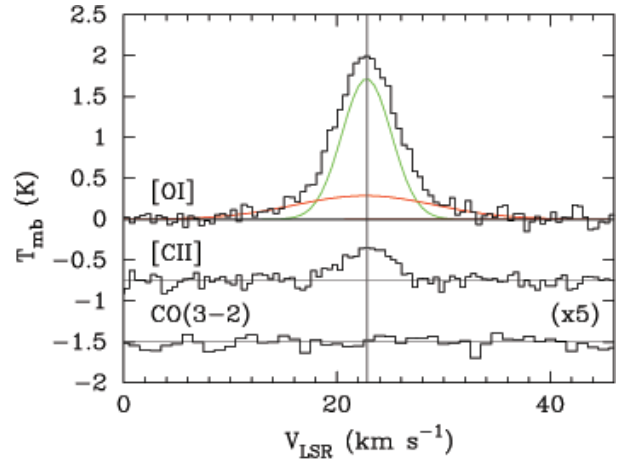}
\figcaption[]{~~CO(3--2), \OI,  and \CII\  spectra towards HD\,50138. ~The
CO(3--2) spectrum was resampled to a velocity resolution of 1 km~s$^{-1}$ and
scaled by a factor of 5 as indicated in the figure (x5). There is no sign of CO emission in the spectrum. The CO(4--3) spectrum,
which is much noisier, see Section 2, is not shown. The \CII\ and \OI\ spectra
were offset in temperature and resampled to a velocity resolution of 0.5
km~s$^{-1}$. The two Gaussian components (Table 1) are plotted  on top of the
\OI\ spectrum in red (broad component) and green (``narrow'' component). The
grey vertical line marks the fitted line center, 22.8 km~s$^{-1}$. All spectra
are labeled. 
\label{fig-spectra1}
 }  
\end{figure}

\begin{figure}[t]
\includegraphics[angle=0,width=8.0cm,angle=0]{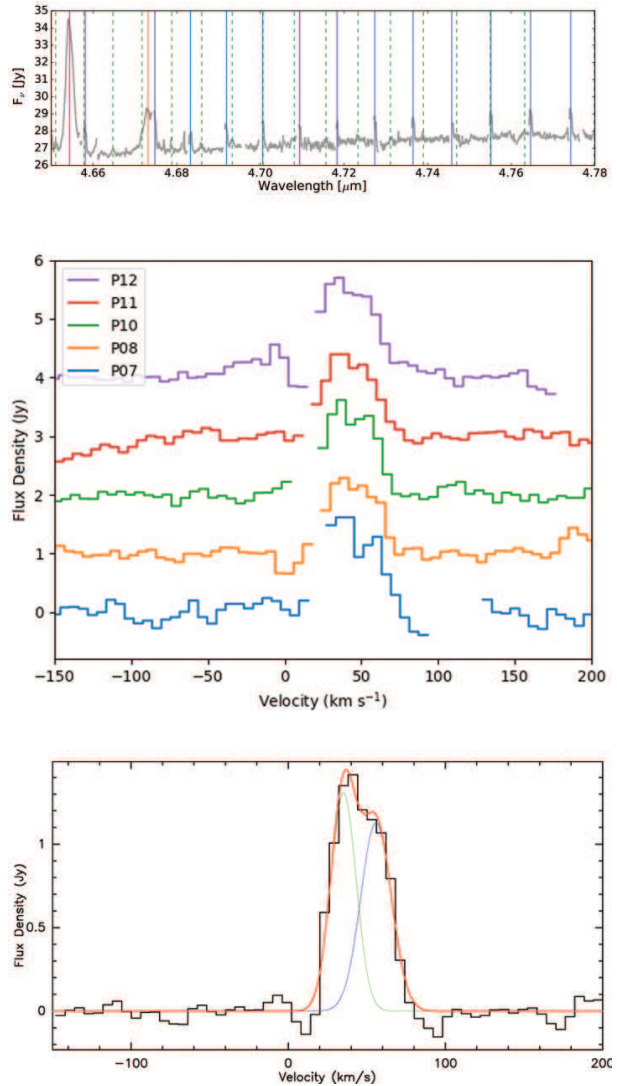}
\figcaption[]{{\it Top panel:} NIRSPEC spectrum covering the wavelength range
4.65 -- 4.78 $\mu$m. The blue vertical lines show $^{12}$CO $\upsilon = 1
\rightarrow 0$ transitions, while the green dashed lines show  $^{13}$CO
$\upsilon = 1 \rightarrow 0$. The red marks the Hu-$\epsilon$ and the purple is
Pf-$\beta$. {\it Middle panel:} Selected CO P transitions from  the NIRSPEC
M-band spectrum. Regions affected by telluric absorption have been blanked
out.The spectra have been continuum subtracted and are offset relative to each
other to improve clarity. {\it Bottom panel:}  Two component Gaussian fit to the
average of all the spectra in the top panel.  The red lines shows the sum of the
two velocity components, while the green and the blue lines show the individual
velocity components. The velocity scale is heliocentric. 
\label{fig-nirspec}
 }  
\end{figure}

Both \OI\ and \CII\ were detected with high signal-to-noise towards HD\,50138
(Fig. 1). The \OI\ line is not well fit with a single gaussian, because it shows
broad faint line wings at low levels, see Table 1. A two component fit gives the
same integrated intensity as a simple integration over the velocity range
covered by the line, 16 K~km~s$^{-1}$. This is somewhat low compared to what we
would expect to see with GREAT from the observed PACS line intensity, 24 $\pm$ 1
$\times$ 10$^{-16}$ W~m$^{-2}$, which corresponds to $\sim$ 21.4 K km~s$^{-1}$
(T$_{mb}$), suggesting that we were probably off source by 1\arcsec\ --
2\arcsec. The high critical density of \OI, 5 $\times 10^5$ cm$^{-3}$ (collision
partner H$_2$, at T$_k$ $\sim$ 120 K, rate coefficient from \citet{Jaquet92}),
requires that the line originates in dense PDR gas. Shock excitation can be
ruled out because there is no evidence for strong shock emission toward the
star. Since the color excess, E$_{B-V}$ toward HD\,50138 is at most 0.15,
with about half of it, 0.08 mag, being interstellar \citep{Borges09}, the
reddening from circumstellar material  is small, $\sim$0.2 mag (assuming an
average galactic extinction law of 3.1), the \OI\ emission cannot originate in a
circumstellar shell. The only plausible explanation is that the \OI\ emission
comes from a disk surrounding the star, especially since the large line width,
6.2 km~s$^{-1}$, and symmetrical line profile are hard to explain except by
rotational broadening in a Keplerian disk. The disk-like morphology of the
circumstellar environment surrounding HD\,50138 is well established both from
polarimetric and spectropolarimetric observations
\citep{Vaidya94,Bjorkman98,Harrington07} as well as from interferometric near-
and mid-IR observations \citep{Monnier09, Borges09, Ellerbroek15, Kluska16,
Lazareff17}. \citet{Ellerbroek15}, who observed HD\,50138 with high spectral and
spatial resolution in the Br$\gamma$ line with AMBER on VLTI, found that they
could  model their data with a thin Keplerian disk and a spherical halo on top
of a Gaussian continuum. Their model with a 0.6 au disk in Keplerian rotation,
with the continuum (hot dust) being more extended, does a relatively good job in
explaining most of their observed features. Additional support for a Keplerian
disk comes from the fundamental CO lines presented in this paper. The shape of 
the broad P-branch lines (Fig.~\ref{fig-nirspec}) suggest that that they would
resolve into clear double-peaked profiles, if observed with higher spectral
resolution. The bottom panel in Fig.~\ref{fig-nirspec} shows that the line
profile can be well fit with a Gaussian with two velocity components; a
blue-shifted one at  +35.1 km~s$^{-1}$ and a red-shifted one at +55.9
km~s$^{-1}$, with line widths of 19 and 23 km~s$^{-1}$, respectively. Both
Br$\gamma$ and CO show asymmetric line profiles, while the \OI\ line appears
perfectly symmetric (Fig.~\ref{fig-spectra1}).

If we adopt the stellar mass,  6 \Msun, a disk inclination, 56\degr\ 
\citep{Ellerbroek15}, and take the rotational velocity as FWHM/2, i.e.,  v$_r$ =
3.1 km~s$^{-1}$ (Table 1), we find that the \OI\ emission originates within a
380 au radius. This sounds  entirely plausible. The broader, faint component
comes from hot gas in the inner part of the disk, where one also sees CO
emission. Although one would naively expect the \OI\ profile to be
double-peaked, the typical signature of a Keplerian disk, there is no evidence
for a central dip in the \OI\ spectrum. This suggests that the \OI\ emission is
optically thin. The same  is  true for well studied bona-fide Keplerian disks
like HD\,100546 and HD\,97048, which also show  single-peaked \OI\ 63 $\mu$m
line profiles (G\"usten et al., 2018, in preparation).

The PACS observations by \citet{Fedele13} show that the \CII\ emission
is extended at the 20\arcsec \ level. Therefore most of the \CII\ emission is
likely to originate in a low density shell created by a past ejection event, although
there could be some contribution to the observed \CII\ emission for the ionized surface layers
of the circumstellar disk. Towards HD\,50138 the
observed  velocity, V$_{\rm LSR}$ = 22.8 km~s$^{-1}$, is the same for both \OI\ and
\CII{}(Table 1), corresponding to a heliocentric velocity, v$_{rad}$ =  40.8
km~s$^{-1}$. This velocity differs somewhat from what has been commonly used,
$\sim$ 35  km~s$^{-1}$ \citep[see e.g.,][]{Borges09,Ellerbroek15}, but is in
good agreement with \citet{Jerabkova16}, RV$_{sys}$ = 40 $\pm$ 4 km~s$^{-1}$.
Since we have much higher velocity resolution than what can be achieved in the
optical, our derived systemic velocity is much more accurate.

\citet{Fedele13}, who did complete range scans of HD\,50138, did not detect any
CO emission, which is often seen in disks around HAEBE stars, see also
\citet{Meeus13}, who analyzed CO transitions from the same data set.
\citet{Fedele13}, however, did detect several OH transitions. From the detected
OH transitions and upper limits to non-detections they derived an excitation
temperature of 130 K and an OH column density of 2 $\times$ 10$^{15}$ cm$^{-2}$.
The relatively cold and low density molecular gas can readily explain why high-J
CO was not detected by {\it Herschel}, but it should be easy to detect in
lower-J CO lines from the ground.  Yet our deep APEX CO(3--2) and (4--3) do not
show any hint of CO emission, nor did \citet{Kama16} detect the  CO(6--5)
transition using the same telescope. If we assume that the excitation
temperature is 100 - 200 K, we find 3-$\sigma$ upper limits of the CO column
density from CO(3--2) of  2.9 $\times$ 10$^{13}$ cm$^{-2}$ and 1.8 $\times$
10$^{13}$ cm$^{-2}$, for an excitation temperature of 100 K and 200 K,
respectively. If the temperature was as low as 50 K, the upper limit from
CO(3--2) would be 3.8 $\times$ 10$^{14}$ cm$^{-2}$. For CO(4--3) the
corresponding upper limits are 6.1 $\times$ 10$^{13}$ cm$^{-2}$ (100~K) and 9.2
$\times$ 10$^{13}$ cm$^{-2}$ (200~K). The CO(6--5) upper limit \citep{Kama16}
provides even fewer constraints to the CO column density, $\sim$ 2
$\times$10$^{14}$ cm$^{-2}$. Therefore for warm (100 - 200 K) molecular gas the
CO column density is $<$ 3 $\times$ 10$^{13}$ cm$^{-2}$. This is unusual,
because CO is always the most abundant molecule in protoplanetary disks (next to
H$_2$, which cannot be observed in the radio regime due to lack of dipole
moment), while OH is  one to several orders of magnitude less abundant
\citep{Bethell09,Visser11}. If we assume that \OI\ comes from the same part of
the disk as OH, i.e., T$_{ex}$ = 130 K,  then our observed line intensity, 11.3
K~km~s$^{-1}$, (Table 1) gives an \OI\ column density,  N(\OI{}) $\sim$ 7
$\times 10^{15}$ cm$^{-2}$, which is about three times larger than the OH column
density, suggesting that the outer disk may have a high fraction of atomic gas.

\begin{figure}[t]
\includegraphics[angle=0,width=8.5cm,angle=0]{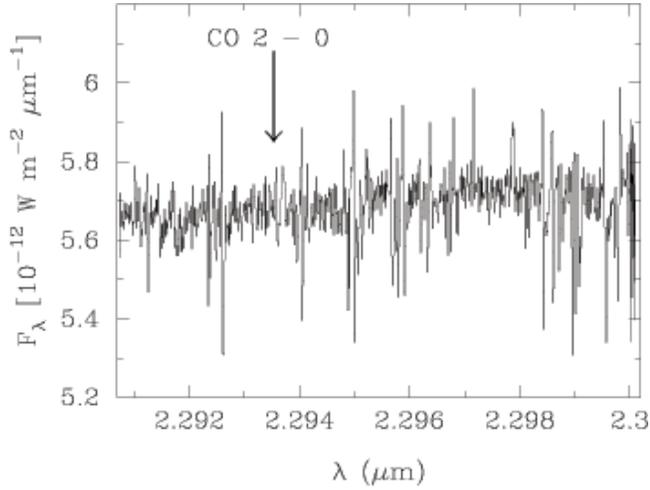}
\figcaption[]{CRIRES spectrum of HD\,50138 covering the wavelength region 2.276 to 2.326 $\mu$m. 
The CO 2 $\to$ 0 bandhead is marked by an arrow. The spikes in the spectra are from imperfect cancellation of telluric lines. 
There is no sign of the CO overtone  bands.
\label{fig-crires}
 }  
\end{figure}



There is no sign of  CO bandhead emission in the  CRIRES spectrum 
(Fig.~\ref{fig-crires}), nor was it seen in a recent X-shooter spectrum
(Benisty, 2017 personal communication). The lack of CO bandhead emission rules
out any hot (2,000 - 3,000 K) CO gas, which is sometimes seen in Herbig Ae/Be
stars \citep{Ilee14,Plas15}, including the unusual B0 star MWC\,349A
\citep{Kraus00}. However, there is clearly warm CO in the inner disk. The
NIRSPEC M-band spectrum (Fig.~\ref{fig-nirspec}) shows that the fundamental
rovibrational $^{12}$CO lines are rather strong. The modeling done by
\citet{Salyk11} give a high column density, N(CO) = 1.3 $\times$ 10$^{14}$
cm$^{-2}$, and a gas temperature of 725 K. The size of the CO emitting region
predicted by their modeling is $\sim$ 2 au, while the size derived from the
observed line width is somewhat larger,  $\sim$ 12 au. This makes the warm CO
too diluted to be detected by {\it Herschel} or any large ground based radio
telescope.

The $^{13}$CO lines appear quite strong relative to $^{12}$CO , see
Fig.~\ref{fig-nirspec}.  We therefore modeled the data to see whether the
$^{13}$CO isotope has a normal interstellar medium (ISM) isotope ratio or
whether it is enhanced.  The model used is a simple slab model \citep[see][for
more details]{Salyk09} assuming a flat disk  that has a $^{12}$CO column density
of 10$^{18}$~cm$^{-2}$, a temperature of 1000 K, and an emitting area of $\pi$
$\times$ (2.45 au)$^2$, or 18.9 au$^2$ --- arbitrarily scaled to match the data.
This model implicitly assumes that the $^{12}$CO and $^{13}$CO are emitted from
the same region, and have the same temperature and emitting area. It seems that
the ISM  $^{12}$C/$^{13}$C ratio, 70,  does not match the data, while a 
$^{12}$C/$^{13}$C ratio of 10 provides a much better fit, suggesting that
$^{13}$CO is enhanced by at least a factor of 5.

Since the PACS beam size at 160 $\mu$m is 11.2\arcsec\ compared to $\sim$
5\arcmin\ for IRAS at 100 $\mu$m, we can compare the IRAS flux densities to
those of PACS. This way one can verify if there is any extended emission
surrounding HD\,50138, which is not seen by PACS. We performed a greybody fit to
the SMA data \citep{Lee16}, the PACS 160 and 70 $\mu$m photometry (this paper),
IRAS, WISE and MSX photometry. This greybody fit gives a dust temperature
of 370 K, a radius of $\sim$ 500 au (not well constrained), and a very small
dust emissivity, 0.12, suggesting large dust grains. This fit is shown in Fig.
~\ref{fig-sedfit}. We can see that the hot dust close to the star starts to
dominate at $\sim$ 10 $\mu$m, which is also seen in the SED plot by
\citet{Lee16}. The color corrected IRAS flux densities agree well within
measurement errors at 25 $\mu$m and 60 $\mu$m with what we derive from our fit;
39.2 Jy and 9.4 Jy at 25 $\mu$m and 60 $\mu$m respectively,  while the  IRAS
flux densities are 44.3  $\pm$  1.9 Jy and 10.5 $\pm$ 0.8 Jy, suggesting that
there is no residual dust cloud around the star. This analysis shows that the
warm dust emission has an angular extent similar to the size we deduce
from \OI. Although the SMA observations did not resolve the dust emission at
1.3~mm \citep{Lee16}, our analysis shows that it could most likely be resolved
by ALMA. It might even be possible to detect warm CO from the inner disk by ALMA
in band 9 or 10.

 

%
%
%
%

\begin{figure}[tbp]
\includegraphics[angle=0,width=8.5cm,angle=0]{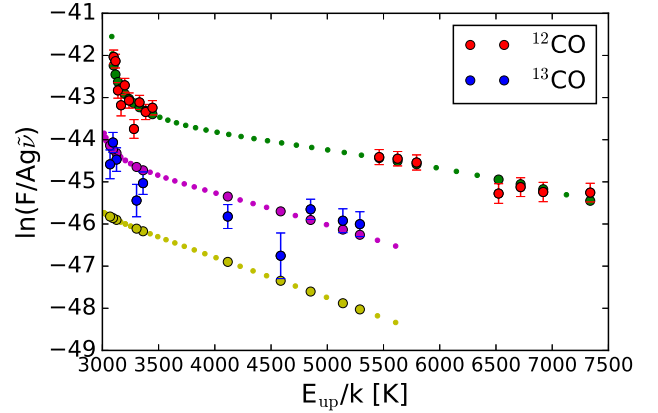}
\figcaption[]{{\bf The y axis is the natural logarithm of flux density, F, 
normalized by the degeneracy wavenumber, Einsteins A-coefficient, and wavenumber
of the transition. Units are in cgs. The $^{12}$CO model (green dots, upper
curve) fits the data quite well. The two $^{13}$CO models shown have a 
$^{12}$C/$^{13}$C ratio of 10 (middle curve, magenta dots) and 70 (lower curve, 
yellow dots).  A $^{12}$C/$^{13}$C ratio of 10 appears to fit the  $^{13}$CO
data reasonably well.}
\label{fig-rot_diagram}       
 }  
\end{figure}


\begin{figure}[t]
\includegraphics[angle=0,width=8.5cm,angle=0]{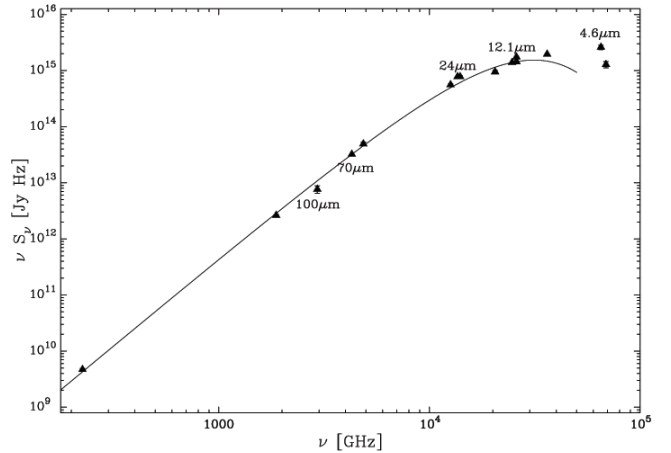}
\figcaption[]{Graybody fit to the millimeter (SMA) and far-infrared (PACS, IRAS, WISE,
and MSX) data of HD\,50138, see text. In the mid-IR, e.g. the MSX band A (8.28
$\mu$m), the hot dust emission from the inner disk is already completely
dominating. Since the x-axis is in frequency, we have labeled some of the data
points in$\mu$m to make it easier to identify the data points discussed in the
text.
\label{fig-sedfit}       
 }  
\end{figure}

\begin{deluxetable}{lrrcc}
\tabletypesize{\scriptsize}
\tablecolumns{5}
\tablenum{1}
\tablewidth{0pt} 
\tablecaption{Gaussian fits to \OI\ and \CII. For CO(4-3) and (3-2), both of which are non-detections, 
we only give the 3-$\sigma$ upper limit of the line integral. Upper limits are computed with a 
4 km s$^{-1}$ velocity resolution over a 20 km s$^{-1}$ velocity range centered on the stellar velocity.
\label{tbl-1}}

\tablehead{
\colhead{Line} & \colhead{$\int T_{\rm mb}^{*} dV$} & \colhead{$T_{\rm mb}^{*}$} & \colhead{$\Delta V$} & \colhead{$V_{\rm LSR}$}  \\
        &   \colhead{[K km s$^{-1}$]}& \colhead{[K]} & \colhead{[km s$^{-1}$]} & \colhead{[km s$^{-1}$]} \\
}
\hline
\startdata
\OI\ &    11.3 $\pm$ 1.2 &  1.7\phantom{0}  & \phantom{0}6.2 $\pm$ 0.3 &  22.78 $\pm$ 0.07 \\
 &       4.7 $\pm$ 1.2 &  0.29 &  15.5 $\pm$ 2.8 &  22.6\phantom{0} $\pm$ 0.7\phantom{0} \\
\CII\   &   2.2   $\pm$ 0.2 &   0.4\phantom{0}&  \phantom{0}5.2 $\pm$ 0.4  &  22.9\phantom{0} $\pm$ 0.2\phantom{0}  \\
CO(4--3)          &     $<$  0.23 &   \nodata &  \nodata &  \nodata  \\
CO(3--2)           & $<$ 0.04 & \nodata & \nodata & \nodata \\
\enddata
\end{deluxetable}

\section{HD\,50138 is not a Herbig Be star}

\citet{Lee16} provided convincing arguments that HD\,50138 is not a HAEBE star.
In particular they could not find any star forming region in the vicinity of the
star from which the star could have been formed. To confirm the isolated nature
of the star we checked the CO images from the all-sky {\it Planck\footnote{
Planck (http://www.esa.int/Planck) is a project of the European Space Agency
(ESA) with instruments provided by two scientific consortia funded by ESA member
states and led by Principal Investigators from France and Italy, telescope
reflectors provided through a collaboration between ESA and a scientific
consortium led and funded by Denmark, and additional contributions from NASA
(USA).}} mission, which provides an unprecedented sensitivity to find molecular
clouds anywhere in our Galaxy. These confirm that HD\,50138 is isolated. The
nearest molecular cloud is $\sim$ 0.5\degr\ to the north. This cloud is
possibly connected to G217.93$-$02.07, which would place it at a distance of
$\sim$ 2 kpc \citep{Elia13}. This distance is much larger than the distance to
HD\,50138, clearly demonstrating that HD\,50138 is not associated with this
cloud. There is a small cold cloud detected at the 5  -- 6 sigma level $\sim$18\arcmin\
west of the star  listed in the {\it Planck} catalogue of Galactic Cold Clumps, 
although it is unlikely that it has any association with HD\,50138.  The second {\it Gaia} release
 \citep{Gaia16, Gaia18} shows that the proper motion is small [(3.6, -3.7) mas/yr] or $\sim$ 8.5 \kms.
The radial velocity is therefore the dominant velocity component. Since the extinction is only 0.4 mag, there is
nothing along the line of sight from which it could have formed. In addition, the PMS lifetime for a 6 \Msun\ star is rather
short, and under that assumption the star could not have wandered very far.
 
It is also well documented that  HD\,50138 has undergone several shell events
\citep{Jerabkova16}, which makes it different from HABE stars, which show no
such activity. The disk around HD\,50138 differs from disks around HAEBE stars.
As shown in this paper there is no detectable molecular gas except for OH in the
extended warm disk where we see \OI. This is very unusual. Searches for CO show
that it is is one to two orders less abundant than OH, while it is the other way
around in protoplanetary disks. We have detected CO, but only in the hot inner
disk. The greybody fit (Section 3), suggests that there is very little cold dust
in the disk. There are certainly HAEBE disks without cold dust like
LkH$\alpha$,101 and MWC\,297. This is shown by detailed SED analysis of
high spatial resolution radio, millimeter and submillimeter emission
\citep{Sandell11}, which confirms that all the emission through the
submillimeter wavelength regime is still completely dominated by thermal
emission from an ionized disk wind.  However, these are all early B-stars with
strong FUV radiation. Cold dust should survive in the mid-plane of a disk around
a B7 star. In this case,  the disk appears dominated by hot ionized gas and warm
atomic gas.

The disk has some unusual characteristics. It appears to be asymmetric
\citep{Ellerbroek15,Kluska16}. Near-infrared interferometric monitoring on the
VLTI \citep{Kluska16} show strong morphological changes in the innermost part of
the disk on a timescale of a few months. \citet{Kluska16} find that they can
reproduce the variability by a model of a disk with a bright spot in the disk,
but they cannot reproduce the variability with a binary model. There is very
little cold dust in the disk and the outer disk appears carbon deficient. It has
OH but no detectable CO, yet the disk is rather extended (720 au). The \OI\
observations show that it has dense warm atomic gas similar or larger than the
amount of OH. However, our modeling of the M-band fundamental CO transitions show
that $^{13}$CO is enriched by more than a factor of five, which excludes it being a PMS star, since such an enhancement 
requires a significant enrichment of $^{13}$C \citep{Kraus09}.

 In short, we agree with \citet{Lee16} that HD\,50138 is not as
pre-main-sequence star. 

\section{Summary and Conclusions}

There is firm evidence that the circumstellar material around HD\,50138 has a
``disk-like'' morphology. Furthermore, \citet{Ellerbroek15}, doing high spectral
and spatial imaging of the Br$\gamma$ line, found that the hot inner disk is in
Keplerian rotation around the central star.  This is also supported by the CO fundamental rovibrational spectra  presented in this paper.
Because of the high critical density
of the \OI\ 63 $\mu$m line, the \OI\ emission must originate in the disk. If we
assume that the outer disk also follows Keplerian rotation, the size of the \OI\
disk is about 760 au based on the observed line width of the \OI\ emission. 
Where \CII\ originates is unclear. Since PACS observations show that the \CII\
emission is extended, most of the emission must
originate in a lower density shell surrounding the star, although we cannot exclude that some of the
emission could come from the ionized surface layers of the disk.

We find that the gas in the extended disk surrounding HD\,50138 is largely ionized and
atomic and the hot dust completely dominates the dust emission. The outer disk has rather unusual chemistry.  \citet{Fedele13} detected several
transitions of OH in their PACS range scan, but no high J CO emission. Long integration 
APEX observations failed to detect CO(3--2) and CO(4--3), indicating that CO  is at
least an order of magnitude less abundant than OH in the extended disk. In
protoplanetary disks CO  is always the most abundant molecule  and OH is
one or several orders of magnitudes less abundant. We find that $^{13}$CO is enriched by more than a factor of five, 
which excludes it being a PMS star, since such an enhancement 
requires a significant enrichment of $^{13}$C in the circumstellar gas as shown by \citet{Kraus09}.

We agree with \citet{Lee16} that HD\,50138 is not a HAEBE star. It must be a main or post main sequence, most likely an FS CMa star,
and we provide further support for the isolated nature of the star. 

Our high spectral resolution \OI\ and \CII\ observations provide a very accurate
radial velocity of the star, 40.8 $\pm$ 0.2 km~s$^{-1}$, i.e., much more
accurate than what can be obtained from optical spectroscopy.
   
\acknowledgements
We thank the anonymous referee for constructive comments, which helped us improve the manuscript.
This research is based on observations made with the NASA/DLR Stratospheric Observatory for
Infrared Astronomy (SOFIA). SOFIA is jointly operated by the Universities Space
Research Association, Inc. (USRA), under NASA contract NAS2-97001, and the
Deutsches SOFIA Institut (DSI) under DLR contract 50 OK 0901 to the University
of Stuttgart.   This work has made use of data from the
European Space Agency (ESA) mission {\it Gaia}
(\url{https://www.cosmos.esa.int/gaia}), processed by the {\it Gaia} Data
Processing and Analysis Consortium (DPAC,
\url{https://www.cosmos.esa.int/web/gaia/dpac/consortium}). Funding for the DPAC
has been provided by national institutions, in particular the institutions
participating in the {\it Gaia} Multilateral Agreement.

%
%

\end{document}